%% file: csfinal.tex
\documentclass{JINST}
\usepackage{graphicx}

\title{Emulsion sheet doublets as interface trackers for the OPERA experiment.}

\input{author.tex}

\author{
\OperaAuthorList
\OperaInstitutes
E-mail: \email{ariga@flab.phys.nagoya-u.ac.jp}
}

\abstract{
New methods for efficient and unambiguous interconnection between electronic counters and target 
units based on nuclear photographic emulsion films have been developed. The application to the OPERA experiment,
that aims at detecting $\nu_\mu \rightleftharpoons \nu_\tau$ oscillations in the CNGS neutrino 
beam, is reported in this paper. In order to reduce background due to latent tracks collected before installation in the detector, on-site large-scale treatments of the emulsions (``refreshing'') have been applied. 
Changeable Sheet (CSd) packages, each made of a doublet of emulsion films, have been designed, assembled and coupled 
to the OPERA target units (``ECC bricks''). A device has been built to print X-ray spots for accurate 
interconnection both within the CSd and between the CSd and the related ECC brick. Sample emulsion 
films have been extensively scanned with state-of-the-art automated optical microscopes. Efficient 
track-matching and powerful background rejection have been achieved in tests with electronically tagged penetrating muons. 
Further improvement of in-doublet film alignment was obtained by matching the pattern of low-energy electron tracks. 
The commissioning of the overall OPERA alignment procedure is in progress.
}

\begin{document}
\section{Introduction}

After decades of successful application in high-energy physics experiments, the nuclear photographic emulsion technique is still attractive for its unique tracking accuracy. Owing to the evolution of fast automated optical microscope systems, it can eventually meet the challenge of ever-increasing scanning areas.

In fact, after the remarkable improvements of the technique in the recent ``hybrid'' experiments\footnote{ $i.e.$ electronic counters to tag and locate neutrino interactions, coupled to a pure emulsion target or to a sandwich metal-emulsion (``ECC'') target to select rare topologies.}  
CHORUS \cite{chorus} at CERN and DONuT \cite{donut}  at Fermilab, new challenges had to be taken up for the OPERA project \cite{opera}. 
OPERA is searching for the appearance of $\nu _\tau$ in the CNGS $\nu _\mu$ beam, a consequence of neutrino oscillation over the long distance separating the source at CERN, Switzerland, from the detector, in the Gran Sasso underground laboratory near L'Aquila, Italy.

In this paper we report the design of interface emulsion film doublets called Changeable Sheets (CSd), suitable for the OPERA purposes, their implementation and the initial performance. 
The basic concepts of the OPERA experiment are recalled in Sec. 2 with a brief overview of the experimental apparatus and of its basic target units, the so-called ECC bricks. 
The motivations to insert CSd units and the strategy for their exploitation are outlined in Sec. 3. 
The details of the large-scale film handling, the CSd packing and initial performance is reported in Sec. 4. 
The alignment methods exploited in experimental tests performed during the commissioning run in 2007 are scrutinized in Sec. 5.
Conclusions and outlook are presented in Sec. 6.

\section{The OPERA experiment}

The OPERA experiment \cite{opera} has been designed to directly validate the flavor-mixing neutrino oscillation model that explains the disappearance of $\nu_\mu$ in atmospheric neutrino experiments \cite{sk}\cite{macro}\cite{soudan} and in accelerator experiments \cite{k2k}\cite{minos}.
 For this purpose the appearance of $ \nu_\tau$ in a pure $\nu_\mu$ beam at a long distance from the neutrino source is searched for. 
 The distinctive feature of $\nu_\tau$ charged-current interactions is the production of a short-lived $\tau$ lepton (c$\tau$  =87 $\mu m$). Thus, one has to accomplish the very difficult task of detecting sub-millimeter $\tau$ decay topologies out of the huge background of $\nu_\mu$ interactions in a target of several $kilotons$. 
This is achieved in OPERA using the nuclear emulsion technique that features an unrivaled 
spatial resolution ($\leq$ 1 $\mu$m). 
The same method was adopted in the DONuT experiment to achieve the first observation of the $\nu_\tau$ \cite{donut}.

The $\nu_\mu$ source is the wide-band CNGS neutrino beam \cite{cngs} with mean energy $\langle E_\nu \rangle=17~GeV$  traveling 730 $km$ through the earth from CERN to the underground Laboratori Nazionali del Gran Sasso (LNGS) where the OPERA detector is placed. In 5 years of CNGS run about 25,000 neutrino interactions will be collected by OPERA, out of which about 12 identified $\nu_\tau$ charge current interactions are expected with a background of less than 1. This number of events is estimated at the nominal value of the oscillation parameter $\Delta m^2 = 2.5\times 10^{-3}~eV^2$ and full mixing \cite{sk}.

\begin{figure}[bthp]
\begin{center}

\includegraphics[width=0.6\linewidth]{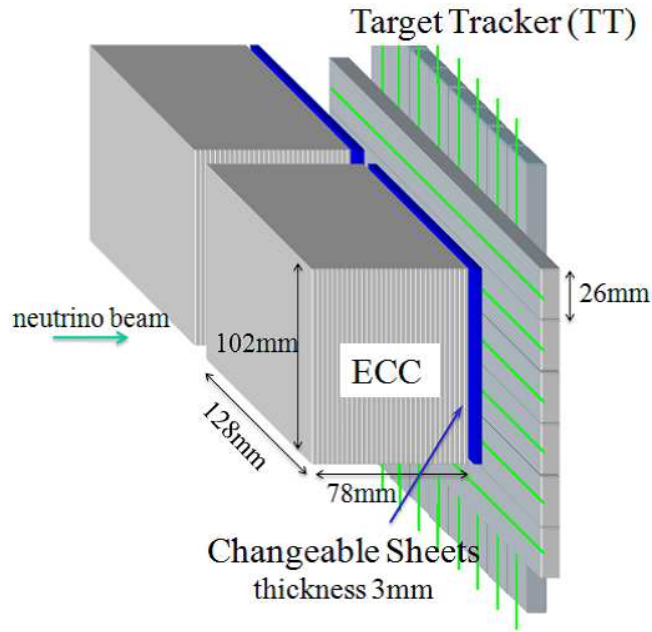}

\caption{The basic elements of the OPERA target section, the ECC brick with its CSd attached, as explained in sec. 3, and the TT.}
\label{fig:ecc-cs-tt}
\end{center}
\end{figure}

OPERA is a large detector ($10~m \times 10~m \times 20~m$) placed in the underground LNGS Hall C 
\cite{run2006}. It consists of two identical super-modules aligned along the CNGS beam direction, each made of a target section and a muon spectrometer. 
Each target section consists of a multi-layer array of 29 target walls interleaved with pairs of planes of plastic scintillator strips. Each plane consists of 256 6.3-$m$ long strips of 26.4 $mm$ $\times$ 10 $mm$ cross-section and measures one of the two transverse coordinates. This constitutes the Target Tracker (TT) \cite{tt}. 
A target wall is an assembly of horizontal trays each loaded with Emulsion Cloud Chamber target units called ECC bricks. Each brick consists of 57 emulsion films interleaved by 56 Pb plates, 1 $mm$ thick, tightly packed and safe-light secured. It has $128~mm \times 102~mm \times 79~mm$ outer dimensions and weighs 8.3 $kg$. Interface emulsion detectors (CSd) are attached to the downstream face of each brick. There are more than 150,000 bricks in total for a target mass of 1.35 $kilotons$. The modular structure of the OPERA target section is sketched in Fig. \ref{fig:ecc-cs-tt}. 

During the exposure to the CNGS beam, data taking, emulsion scanning and analysis are required to progress in parallel. 
Event-by-event, first of all hit reconstruction and $\mu$ tagging are performed based on triggered electronic signals in the TT and the muon spectrometers. 
Then, one or more selected bricks, candidates to contain the latent image of a neutrino interaction, are extracted from the apparatus. The corresponding CSd is first scanned. The brick which contains neutrino interaction is brought up to the surface laboratory, exposed to a moderate flux of cosmic-rays ($\sim$100 $tracks/cm^2$) for the purpose of film-to-film alignment and disassembled. 
Emulsion films therein are chemically developed and made ready for optical scanning.
 
Automated microscopes, available at several OPERA institutions in Japan and Europe \cite{suts}\cite{ess}, allow performing the second emulsion-based data taking. 
This is a complex, multi-step task consisting of the location of a neutrino interaction vertex by scanning backwards (scan-back) along the path of the selected tracks, the vertex reconstruction by scanning a volume of emulsion around the presumed end points of the tracks, the identification and validation of any $\tau$ decay topology and further optional measurements for efficiency study and kinematical analysis.

\section{Changeable Sheets for OPERA}
\label{sec:csforopera}
In the procedure described above, the connection of track elements from the TT to the ECC bricks and to the parent vertex are the most critical issues, because one has to pick-up a small unit in a massive apparatus and make the bridge between the approximately 1 $cm$ transverse resolution of the TT to the 1 $\mu m$ 3D spatial resolution of nuclear emulsions.
For this reason, specially treated nuclear emulsion films, called Changeable Sheets (CS), are used as interfaces between each TT plane and the corresponding ECC bricks \cite{csspsc}.  
CSd stay attached to the downstream face of the bricks all the time during the exposure 
to record beam-related signals. However, upon brick selection and extraction, they are detached and developed underground without any exposure to cosmic-rays. 

The technique of interface emulsion sheets has already been applied in other hybrid experiments \cite{chorus}\cite{donut}. However, special developments were needed for OPERA in order to obtain a very low background and to operate on a very large scale.

The CS play two major roles. The first one is to confirm that the ECC brick which contains the neutrino interaction vertex is the one pointed to by the TT reconstruction. 
The second one is to provide the neutrino-related tracks for the ECC brick scan-back analysis.

Brick confirmation is required in the OPERA analysis given the resolution of the TT and the consequent sizable probability to misidentify the ECC brick (e.g. $\sim$30\% for $\nu_\tau$ charged-current interactions). 
The information obtained after the CS scanning is highly beneficial to avoid useless film handling and processing of the misidentified bricks and minimize the corresponding waste of target mass, also to avoid useless scanning overload of the ECC bricks.
Thanks to the CS the bricks wrongly identified by the TT are not dismantled but put back in the target with a fresh CS attached to them. 
Moreover, where the TT reconstruction is compatible with two or more bricks, these are ordered by probability and their CS are scrutinized accordingly. This significantly increases the event finding efficiency.

As described in the previous Section, the ECC brick is exposed to cosmic-rays. Consequently, there will be a few tracks in the brick related to the neutrino interaction and a large number of cosmic-ray tracks ($\sim10,000~tracks/brick$). 
The CS provides a clear identification of the neutrino-related tracks to be searched in the most downstream film of the corresponding ECC brick as a clean start-up of the scan-back procedure aiming at vertex location. 
In order to recognize genuine neutrino-related tracks and allow brick confirmation, the background in the CS must be suppressed down to extremely low levels.

\begin{figure}[bthp]
\begin{center}
\includegraphics[height=0.6\linewidth]{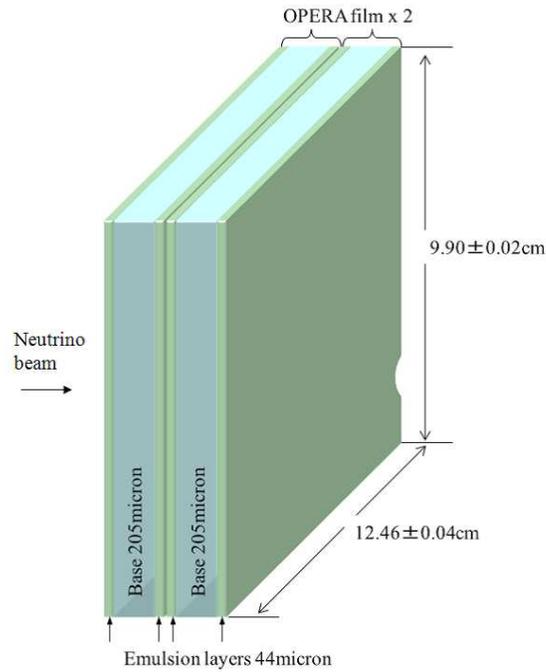}
\caption {The schematic drawing of the CS doublet.}
\label{fig:cs_schematic}
\end{center}
\end{figure}

\begin{figure}[bthp]
\begin{center}
\includegraphics[width=0.6\linewidth]{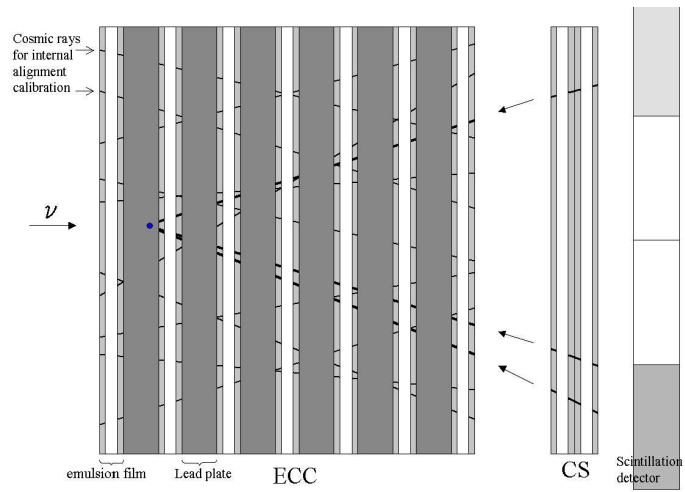}
\caption {Schematics of the CS - Scan-back method.}
\label{fig:scanback}
\end{center}
\end{figure}

The final choice of CS geometry for OPERA has been to assemble two adjacent emulsion films as a $doublet$ (CSd), coupled as an independent, detachable package to the downstream face of an ECC brick as shown in Fig. \ref{fig:cs_schematic}. A film has two emulsion layers. The request of a coincidence between two emulsion films (more than any of 3 recognized track segments out of 4 measured layers) results in a strong background reduction and a high tracking efficiency. 
The overall configuration of a brick unit and a schematic illustration of the OPERA scan-back method for event location is shown in Fig. \ref{fig:scanback}.


Since the background level is due to single-film track density and their fake coincidences, keeping it low requires: 
a) to reduce the single-film physics background due to the continuous sensitivity of the emulsion films\footnote{ All the time from the film production till the chemical development, any ionizing radiation 
induces persistent photographic recording of latent tracks.};  
b) to pack very precisely the two films in a doublet; 
c) to exploit extremely accurate alignment tools and methods. 
What is especially challenging with OPERA is to do this on a very large scale for 150,000 bricks.

\section{Emulsion film handling and CSd production}

\subsection{Film background reduction}

Custom emulsion films jointly developed by the Nagoya University group and Fuji Film Corporation, called $OPERA~ films $ \cite{emul}, are employed both for ECC bricks and CS.
Mass-producible OPERA films consist of two 44 $\mu m$ thick emulsion layers sensitive to minimum ionizing particles placed on both sides of 205 $\mu m$ thick plastic base. 
Their size is $12.46~\pm~0.04~cm$ $\times$ $9.90~\pm~0.02~cm$ at 40\% relative humidity (RH)\footnote{ The expansion factor as a function of RH is 1.4$\times10^{-4}$ /\%.}. 
Film composition (small, regular, finely dispersed sensitive $Ag Br$ crystals) and geometry allow in principle a position resolution of 0.05 $\mu m$ and an angular resolution of 0.4 $mrad$  \cite{angleresolution}. 

A procedure to erase to a large extent any previously recorded latent-image track due to ionization
was developed for the OPERA films. Such a $memory~reset$ called ``refresh'' is obtained by keeping the films at high humidity (95-99\% RH) and high temperature (25-30{$^\circ$C}) for a few days \cite{emul}.

Films produced for the OPERA experiment in the Fuji facility in Japan were first transported to the Tono mine Refresh Facility (Gifu, Japan) to erase the cosmic-ray tracks earlier accumulated. 
Afterwards, the refreshed films temporary packed as brick-wise slots were shipped to LNGS. Hence, additional cosmic-ray background ($\sim$1400 $tracks/cm^2$) was accumulated during the 1 month journey from Japan to Italy. 
In the case of ECC bricks, these background tracks are easily identified and rejected as compliant with a ``travel alignment''\footnote{ The hits made by this cosmic-ray background form a track only under the assumption that the films are not interleaved with lead plates.}.


In the case of the CSd, the films are assembled at random and disassembled underground at the LNGS where the cosmic-ray flux is very low. Thus, the residual background is mainly due to accidental coincidences between early tracks on each film.
They could still spoil the required unambiguous ECC brick confirmation because they compete with the average flux of beam-related tracks expected to be extremely small (e.g. in $\nu_\tau$ charged-current interactions about 4 tracks are expected), although the scanning area of the CSd is extremely large ($\sim$100 $cm^2$). Thus, the background density must be kept well below 10$^{-3}$  $tracks/cm^2$ for $|tan\theta|<0.5~rad$ with $\theta$ being the angle with respect to the normal to the film plane. This requirement is much stronger than in the ECC films (100 $tracks/cm^2$).

As
\begin{equation}
N_{Background} \propto (Single~film~track~density)^2\times (Alignment~accuracy)^2
\label{eq:bg}
\end{equation}
constraints are very demanding on both factors.

In order to reduce the track density in the emulsion films, a second refreshing is performed underground at LNGS to erase tracks accumulated during transportation.
Observed track densities before and after this process are listed in Table \ref{table:bgon1film}. 
The single-film track density after refreshing is below 100 $tracks/cm^2$. 

\begin{table}[h]
\begin{center}
\begin{tabular}{cccc}
\hline
&\multicolumn{3}{c}{Track density $[tracks/cm^2]$}	\\ 
			&Total			&Cosmic-ray		& Radioactivity \\
\hline
Non-Refreshed	&$1410 \pm 125 $	&$870 \pm 95 $	& $ 540 \pm 80 $ \\
Refreshed		&$76 \pm 9 $		&$37 \pm 6 $	& $ 39 \pm 6 $	 \\
\hline
\end{tabular}
\caption{Track density on single CSd emulsion films. Scanning was done by the UTS automated scanning system at the Nagoya University in a standard configuration (95\% detection efficiency for $10~GeV~\pi ^-$ beam tracks). Only tracks recognized on both emulsion layers are accounted for. Tracks were topologically classified  as due to cosmic-rays (very straight penetrating tracks) or radioactivity.}
\label{table:bgon1film}
\end{center}
\end{table}

\subsection{Overview of the CSd production}

The CSd film handling, production and development are hosted in a specially designed
facility situated in the underground Hall-B of LNGS.
The production facility arranged to produce 1,000 $CSd/day$ is partitioned in three rooms. The first two are safe-light darkrooms assigned to refreshing and packing. 
The last steps (envelope folding and insertion in a rigid plastic box) are performed in the third room. 
Five operators may simultaneously work in this facility.

\subsubsection{The refreshing apparatus}

\begin{figure}[bthp]
\begin{center}
\includegraphics[width=0.5\linewidth]{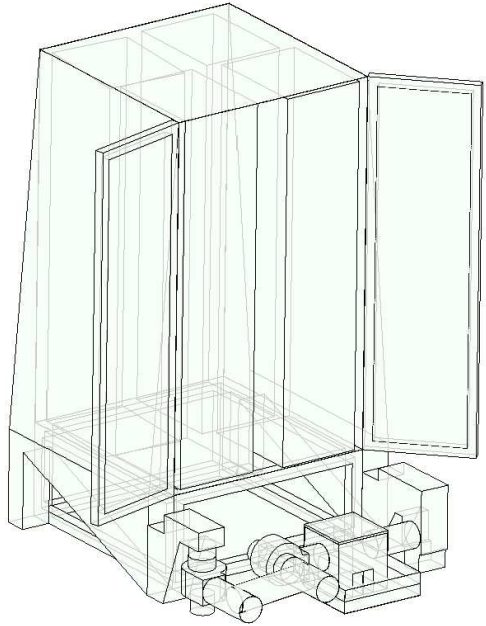}
\caption {A schematic drawing of the refresh chamber. }
\label{fig:chamber}
\end{center}
\end{figure}

In the refreshing room there are two refresh chambers. Each chamber (Fig. \ref{fig:chamber}) $\sim 1~m\times 1~m\times 2~m$, contains 760 trays. Each tray contains 9 films.
In total 6840 films are processed at the same time. 
The component in front of the chamber is a two-fan humidifier which provides humidity through wet paper filters. Demineralized water is used. 
The ducts have tapered structures to provide uniform ventilation inside the chamber. 
The air flow can be in open circuit (air from outside, evacuated outside) or in closed circuit 
(air circulation without exchange).

The air in the refreshing room is kept at 26-27${^\circ C}$ and 40\% RH. A refreshing cycle 
takes one week. The sequence is the following: open circuit mode at $\geq$85\% RH for one day, 
close circuit mode at $\simeq$98\% RH for three days, slow-dry mode from 98\% to 60\% RH for 
one day, complete-dry mode down to 40\% RH for one day (open circuit).

\subsubsection{Packing, production and quality control.}

\begin{figure}[bthp]
\begin{center}
\includegraphics[width=0.7\linewidth]{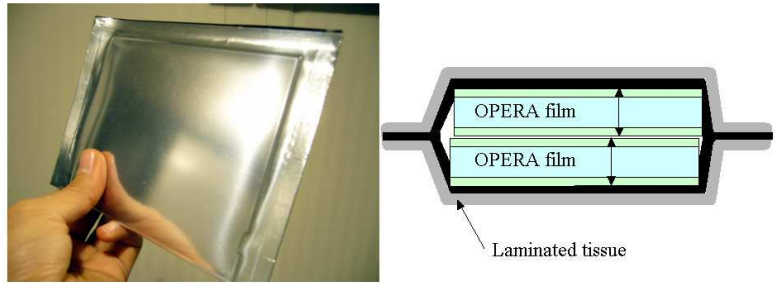}
\caption {The packed CSd and a schematic drawing of the cross-section.}
\label{fig:package}
\end{center}
\end{figure}

The two emulsion films of a CSd are packed together by using an aluminum-laminated tissue. The structure of the laminated tissue is (from outside to inside): /Nylon 15 $\mu m$/ Polyethylene 13 $\mu m$/ Aluminum 7 $\mu m$/ Polyethylene 13 $\mu m$/ Carbon-mixed-polyethylene 35 $\mu m$/. The light tightness is provided by the aluminum layer 
and by the Carbon-mixed-polyethylene layer. The mechanical strength is given by the nylon layer. 
The laminated tissue is shaped as an envelope. The CSd is temporarily vacuum packed (Fig. \ref{fig:package}). 
Sealing is done by welding the polyethylene.

A semi-automated packing machine was developed for mass production. 
A small size (230 $mm$$\times$280 $mm$$\times$1.2 $mm$) vacuum chamber was designed to evacuate air rapidly. It reaches a vacuum of 15-35 mbar within 3 seconds. 
A complete packing cycle takes 10 seconds. With each of the two machines available in the facility an operator can fabricate 850 $CSd/day$. 
The net size of the packed CSd is 125.5 $\pm$ 0.3 $mm$ $\times $100.6 $\pm$ 0.3 $mm$ $\times $1.6 $\pm$ 0.1 $mm$. 
The average displacement between two films is $\langle|\delta x|\rangle$ = 310 $\mu m$, $\langle|\delta y|\rangle$ = 90 $\mu m$.

\begin{figure}[bthp]
\begin{center}
\includegraphics[width=0.7\linewidth]{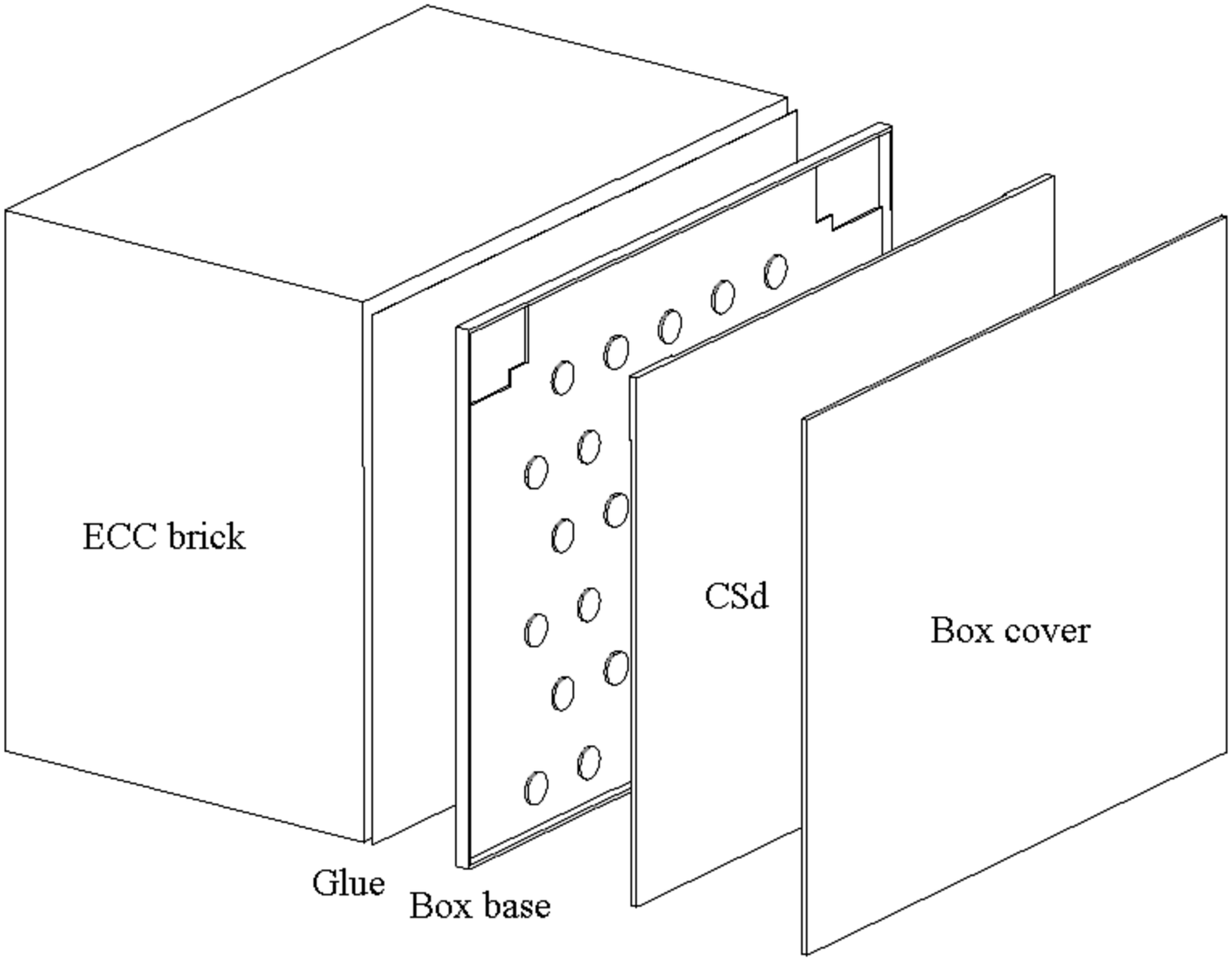}
\caption {Mounting the CSd on the ECC brick.}
\label{fig:ecc-cs}
\end{center}
\end{figure}

The packed CSd is inserted into a specially shaped plastic box and 
then attached to the ECC brick. A schematics is given in Fig. \ref{fig:ecc-cs}. This box renders 
the CSd $changeable$. The outer size of the box is 
127.8 $\pm$ 0.2 $mm$ $\times$ 102.7 $\pm$ 0.2 $mm$ $\times $3.0 $\pm$ 0.1 $mm$. 
The inner size is 126.7 $\pm$ 0.2 $mm$ $\times$ 101.0 $\pm$ 0.2 $mm$ $\times $1.7 $\pm$ 0.1 $mm$. The
CSd envelope tightly fits into the plastic box.

The production of the CSd for the OPERA experiment has started on November 2006 and it is presently
going on (April 2008). After an initial setting-up period 60\% of the total was constantly produced during the last 9 months of 2007.

\begin{figure}[bthp]
\begin{center}
\includegraphics[width=0.7\linewidth]{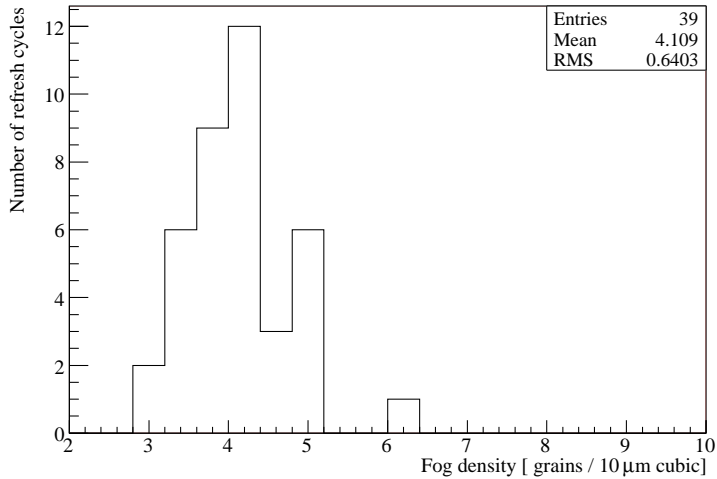}
\caption {The fog density after film refreshing at LNGS.}
\label{fig:refresh_fog}
\end{center}
\end{figure}

During production a regular sampling was performed for film quality control and long-term
CSd quality control. First, it was checked that film sensitivity is fully kept after 
refreshing. Secondly, the number of randomly diffused grains (known as fog density) was monitored cycle-by-cycle. In fact, while tracks are erased by refreshing the fog density is known to increase slightly. An exceedingly high fog density may affect the scanning performance of the automated microscopes. Fog density data are reported in Fig. \ref{fig:refresh_fog}. A sample film was picked-up from each refreshing cycle. Entries of the plot correspond to the first half of the production. 
The initial fog density is well below the safe limit of 10 $grains~ per~ (10 \times 10 \times 10~ \mu m^3)$.

Long-term quality controls were performed on CSd packed under vacuum. It appeared that some gas was produced inside the envelope causing an unacceptable long-term fog increase. In order to avoid that, a pin hole was bored on the edge of each CSd envelope thus allowing the gas to exhaust. The cause of the degassing and the nature of the gas are under investigation.

\begin{figure}[bthp]
	\begin{center}
	\includegraphics[width=0.45\linewidth]{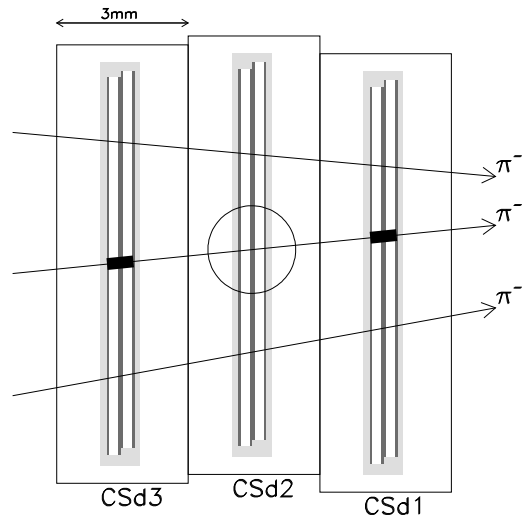}
	\caption { A schematic drawing of the module exposed at CERN. 
                  The tracking efficiency of the CSd was evaluated by checking the presence 
                  in CSd2 of tracks reconstructed using CSd1 and CSd3.}
	\label{fig:cern_module}
	\end{center}
\end{figure}

\begin{figure}[bthp]
	\begin{center}
	\includegraphics[width=0.6\linewidth]{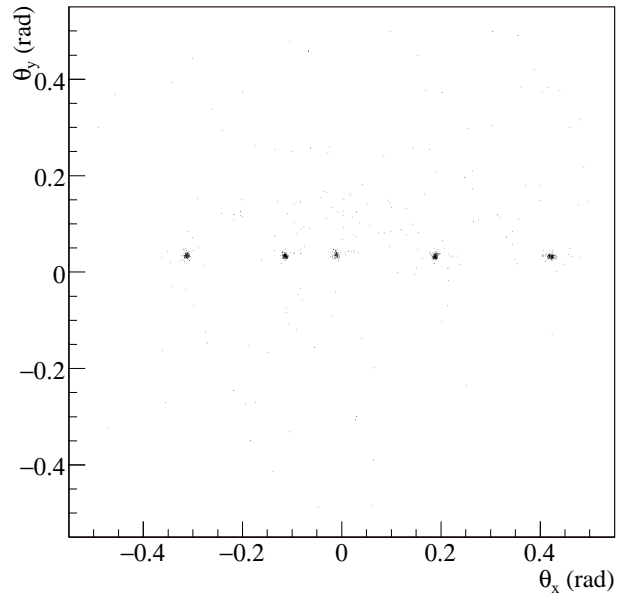}
	\caption { The angular distribution of the $\pi ^-$ measured by the CSd.}
	\label{fig:cern_angle}
	\end{center}
\end{figure}

The performance of the designed CSd was studied by several experimental tests. For efficiency studies a test module was exposed to 10 $GeV$ $\pi ^-$ in the CERN PS T7 beam line. A schematic drawing of the module is given in Fig. \ref{fig:cern_module}. The CSd have been produced at the LNGS facility. Data have been taken in the emulsions at five different angles with respect to the emulsion films spaced by about 0.1 $rad$. 
The angular distribution of the $\pi ^-$ beam is shown in Fig. \ref{fig:cern_angle}. 

\begin{figure}[bthp] \begin{center}
\includegraphics[width=0.7\linewidth]{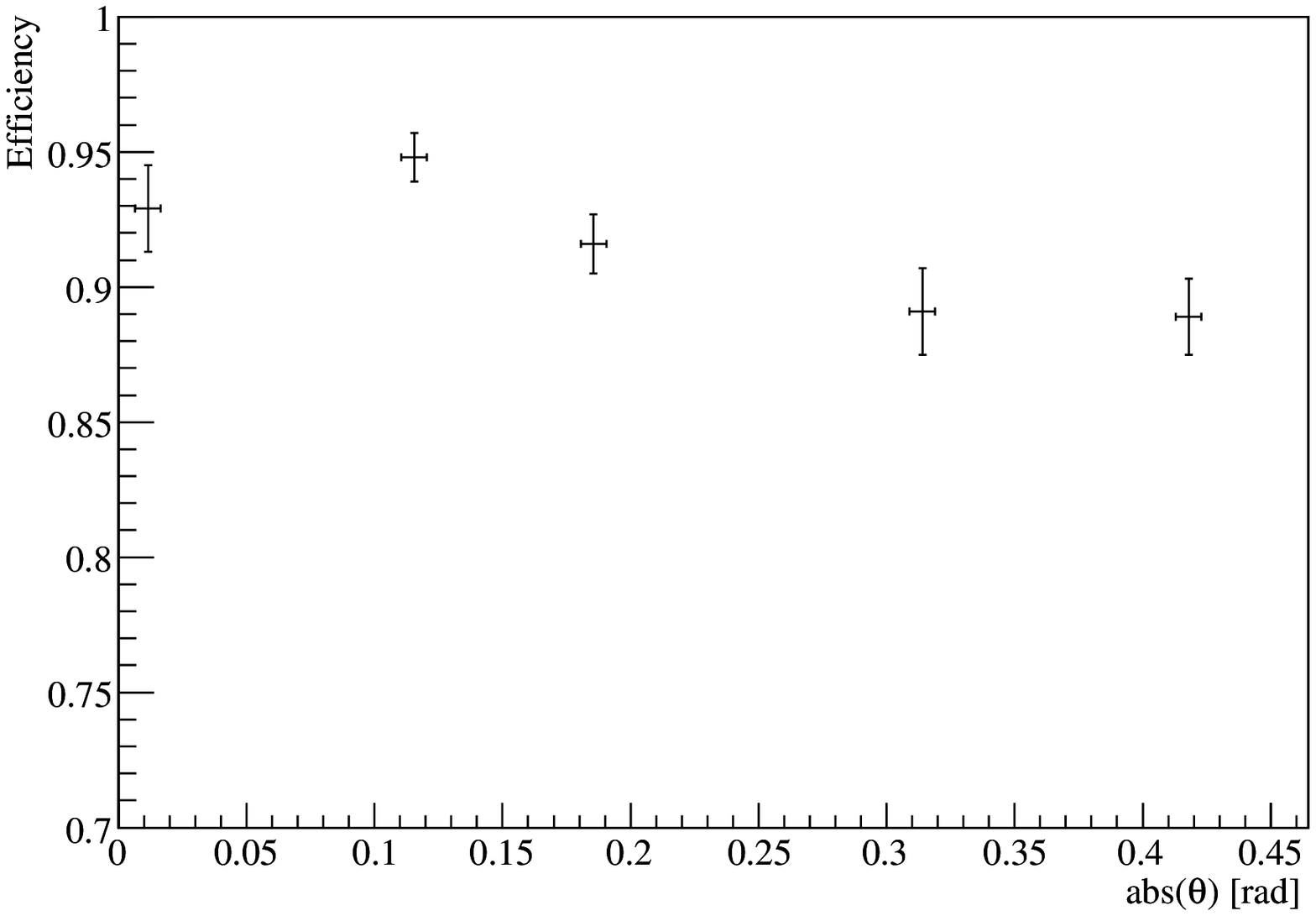}
\end{center}
\caption { Tracking efficiency of the CSd as a function of the angle of the track with respect to the normal to the emulsion film.}
\label{fig:eff}
\end{figure}

Emulsion scanning was performed by using fast automated microscopes at Nagoya University. The three CSd were aligned by using the $\pi^-$ tracks themselves. The tracking efficiency was calculated by reconstructing tracks using CSd1 and CSd3 and then checking the presence of tracks on CSd2. Fig. \ref{fig:eff} shows the tracking 
efficiency and its dependence on the angle. In each CSd a track is defined by a coincidence of four track segments in the four emulsion layers.

\section{Alignment issues}

Four steps will be considered for the OPERA analysis: 
1) TT-to-CSd connection; 2) CSd film-to-film connection; 3) CSd-to-ECC brick connection; 4) ECC brick film-to-film connection.

Alignment is commonly based on passing-through high energy charged particles ($\geq 1GeV$) such as in the test beam exposure reported in the previous section. 
If the exposure density is sufficiently high ($\simeq 10^2~tracks/cm^2$), adjacent emulsion films can be aligned with an accuracy of $\leq 1\mu m$ within a sufficiently small area ($\leq 5~mm\times 5~mm$). 
However, in the OPERA case the high-energy particle flux is extremely low at the LNGS underground laboratory. 
Only step 4 can be based on an exposure to cosmic-rays at the LNGS surface before development. 
Unambiguous connection at step 3 is needed. 
For step 2 even after on-site film refreshing accurate film-to-film connection in the doublet is nonetheless required.

\subsection{Connection between TT and CSd}\label{par:TTCS}  


Data acquisition with cosmic-rays was performed from April 12$^{th}$ to May 16$^{th}$ 2007.  
The number of ECC bricks (with CSd) installed at that time was about 10,000. 
At the end of the run 579 tracks were reconstructed crossing 1203 ECC bricks. 
Muon tracks were further selected according to the following criteria: 

\begin{itemize} 	
\item A slope smaller than 0.550 $rad$ with respect to $Z$, the horizontal projection of the beam direction, to which the emulsion films are orthogonal. The automatic scanning systems have maximum efficiency for tracks perpendicular to the emulsion films, $i.e.$ close to the beam direction. 
Most of cosmic-ray tracks are useless, being nearly vertical. The smallest slope with respect to $Z$ in the sample was 0.350 $rad$.
\item A minimum number of 28 hits in the TT in order to have a long and clean track. 
\item A reconstructed impact point at more than 1$cm$ from the emulsion film edge to ensure that the prediction falls in the CSd fiducial area\footnote{ There is a few $mm$ space between a CSd and the neighbor CSd.}.
\end{itemize}  

\begin{figure} 	
\centering 		
\includegraphics[width=0.45\textwidth]{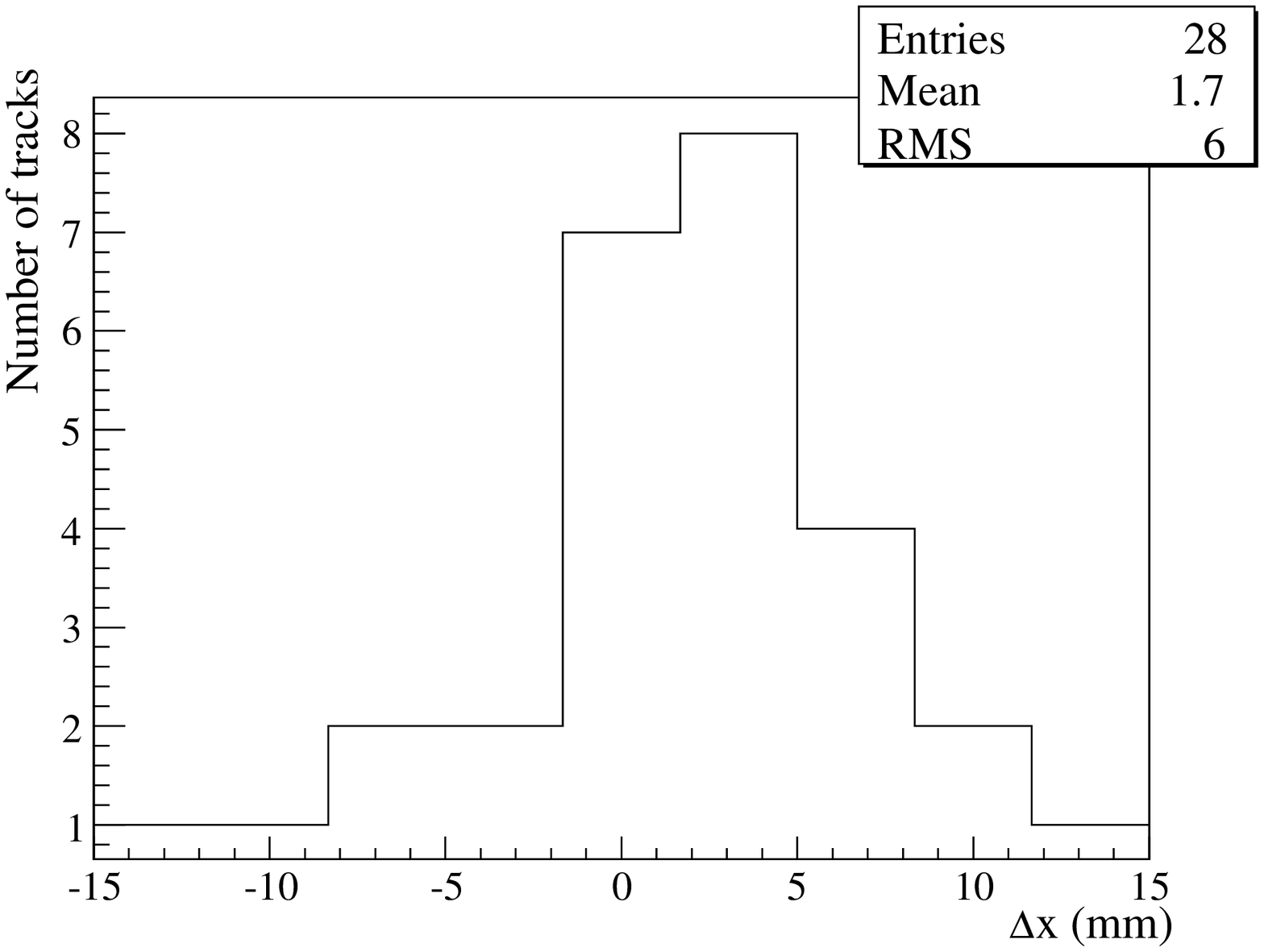} 		
\includegraphics[width=0.45\textwidth]{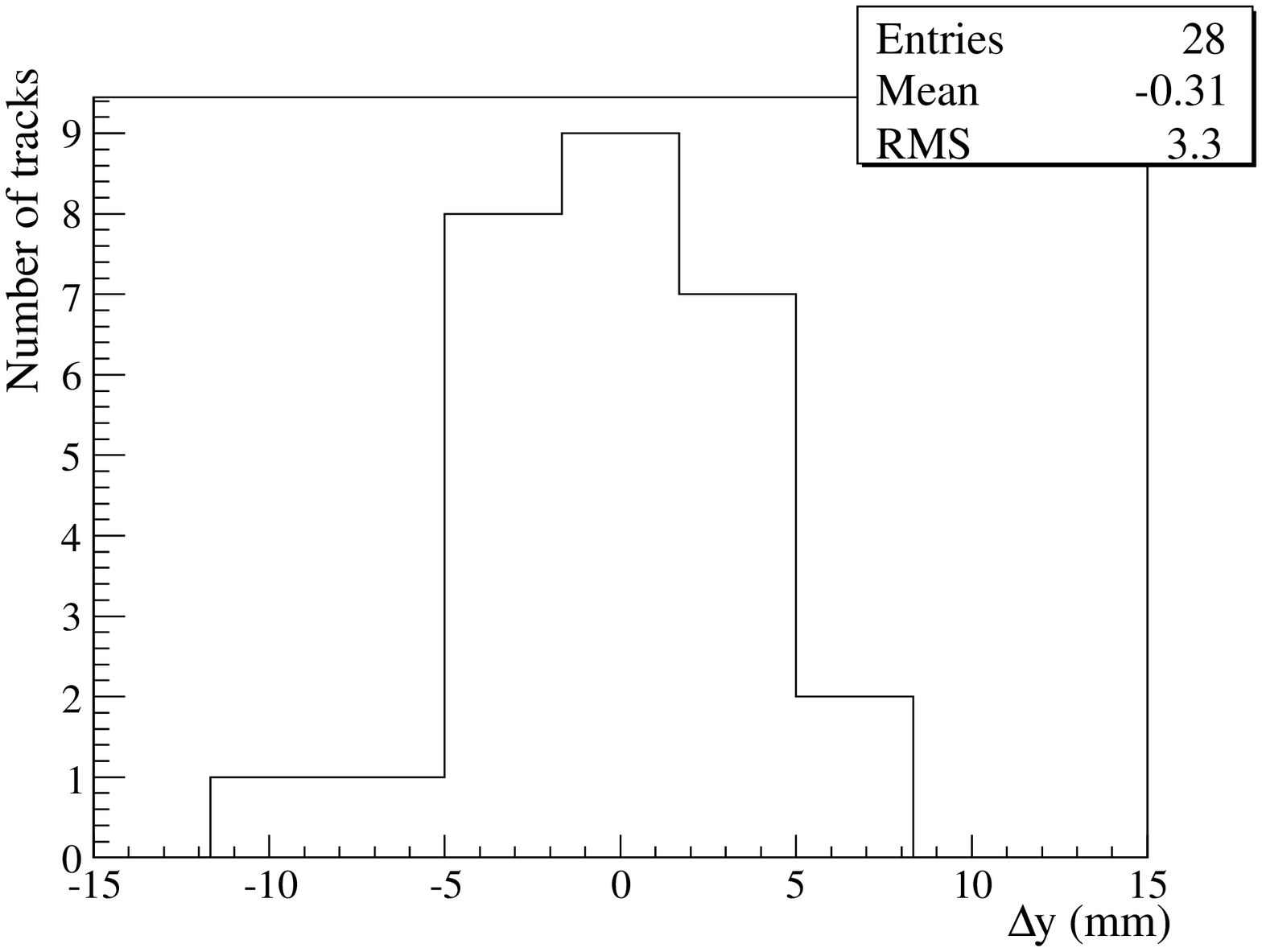}\\ 		
\includegraphics[width=0.45\textwidth]{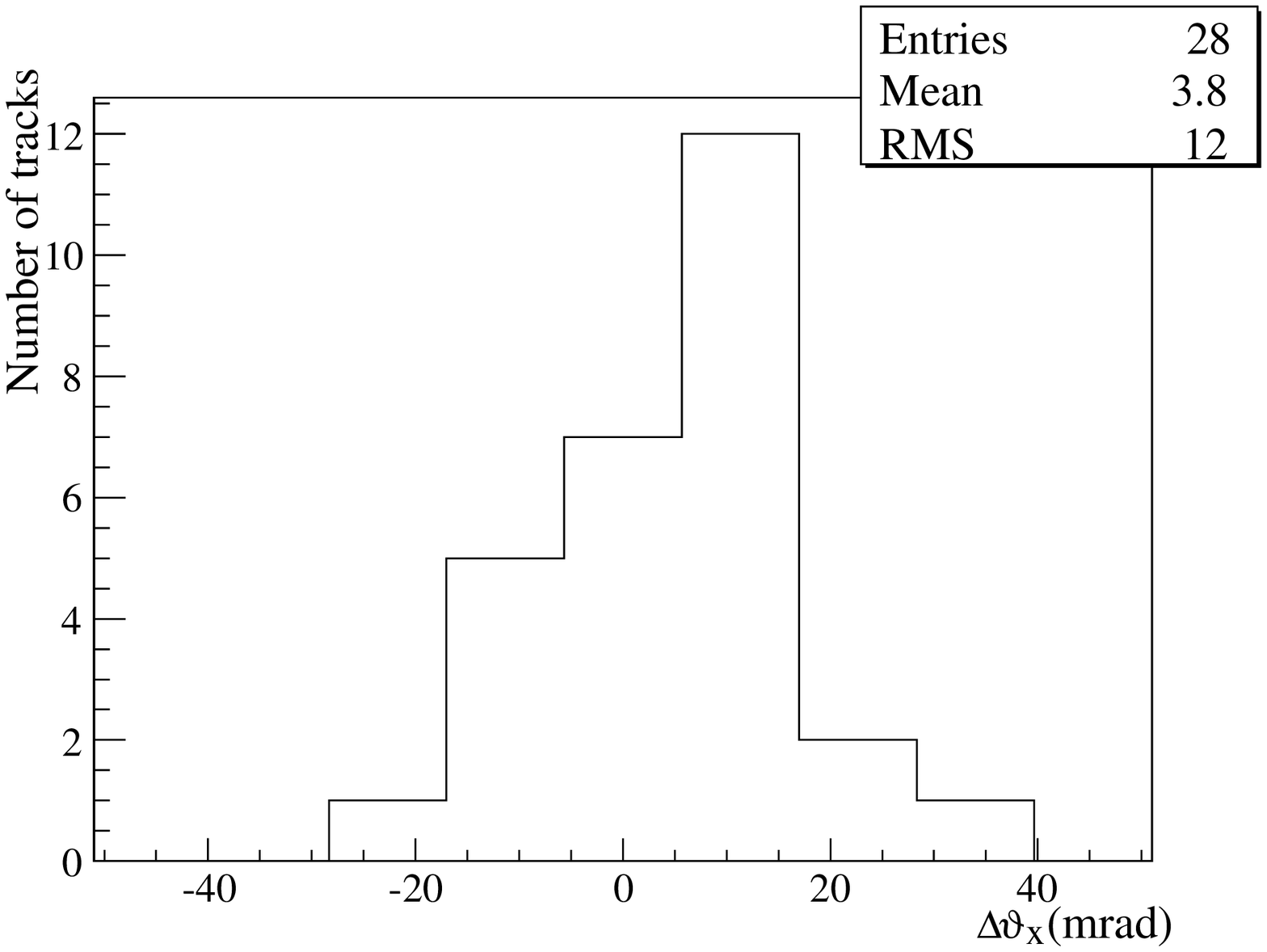} 		
\includegraphics[width=0.45\textwidth]{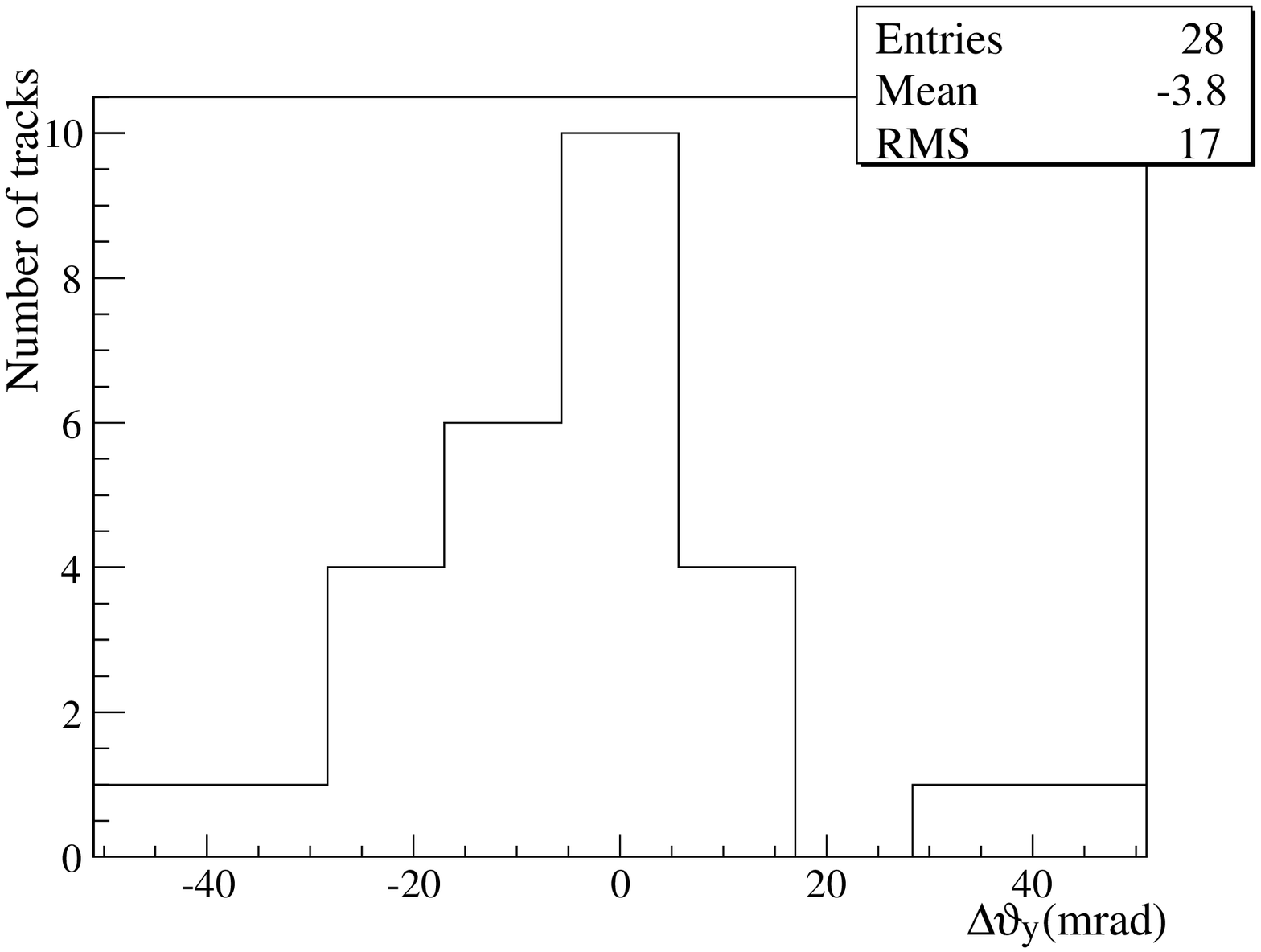} 	
\caption{\small Distributions of the position and angular residuals of the 
emulsion measurements with respect to the TT reconstruction.} 	
\label{fig:TTCS_dist} 
\end{figure}

30 bricks were selected corresponding to 21 events, since a cosmic-ray muon can cross more than one brick. The bricks were extracted, X-ray marked as explained later and the associated CSd was developed in the underground CS facility. The scanning was performed at the Scanning Station at LNGS with the European Scanning System \cite{ess}.  An area of $5 \times 3~cm^2$ centered around the prediction was measured on both films. 
For each CSd film, track segments (hereafter called micro-tracks) were measured in each of the two emulsion layers. Then, two micro-tracks on both sides of the film plastic base were associated to form a line (hereafter called base-track) by using the points where each of the two micro-tracks intercept the nearest surface of the plastic base. 
Contrary to the micro-track the slope of a base-track is largely unaffected by distortions caused to the emulsion layer during film processing.
A CSd track candidate is defined by the matching of either both base-tracks, thus all 4 micro-tracks, or a base-track and a single micro-track, thus 3 micro-tracks. 28 muon track candidates out of 30 predictions were located in the CSd scanning. Out of those, 23 had all 4 micro-tracks found. 
Fig. \ref{fig:TTCS_dist} shows the distributions of the position and angular residuals of the emulsion measurement with respect to the TT reconstruction for both X and Y transverse projections.  

\subsection{CSd film-to-film connection}

The film-to-film-connection inside the CSd is a key alignment issue to reduce the CSd background (Eq. \ref{eq:bg}). It has been performed by two different methods and was experimentally tested by using the samples of events from CNGS commissioning runs in 2007.

The first method is based on circular X-ray marks of about 100 $\mu m$  diameter printed on the four corners, penetrating both emulsion films in the CSd and the most downstream film of the ECC brick. This provides a common, accurate reference frame for scanning of those 3 films.

\begin{figure}[htbp] 	
\centering 		
\includegraphics[width=0.45\textwidth]{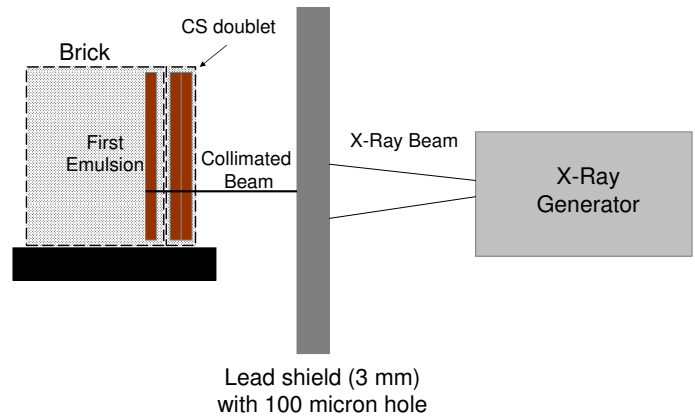} 	
\caption{Schematic view of the X-ray system.} 	
\label{fig:SchemaXray2} 
\end{figure}

\begin{figure}[htbp] 	
\centering 		
\includegraphics[width=0.50\textwidth]{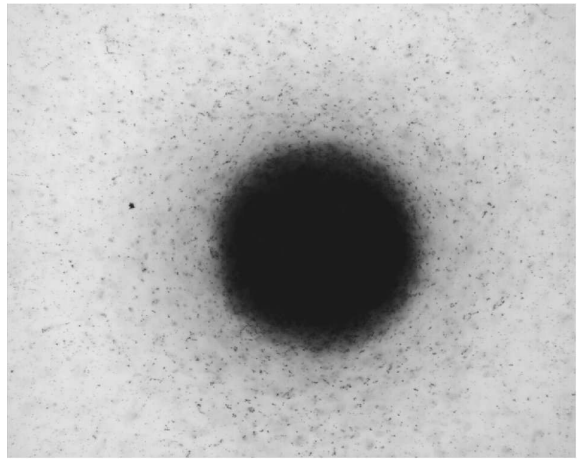} 	
\caption{X-ray spot on an emulsion sheet ($\sim$100 $\mu m$).} 	
\label{fig:mark} \end{figure}

A custom system for X-ray marking of emulsion has been realized (Fig. \ref{fig:SchemaXray2}). 
It is based on a small X-ray generator capable to produce a beam in the range 50 $kV$  to 80 $kV$   with a 0.1 to 2s duration. The beam is collimated through a 100-$\mu m$  hole in a 3-$mm$ thick lead shield. A spot observed in the emulsion is shown in Fig. \ref{fig:mark}. 
Accurate positioning of $\simeq 5 \mu m$ on both axes of the brick with respect to the collimated beam is obtained using a motorized stage.
A safe lead box shielding hosts the system. An user-friendly software interface was 
developed to allow the operator to move the bricks and control the shutter of the X-ray generator.
The complete sequence to print the 4 X-ray spots can be done in a fully automatic mode 
with on-line safety sensor checks. 
  
However, for films in contact, the accuracy of the X-ray mark method is limited by any local deformations of the emulsion films that could amount to about $1~\mu m/cm$. Thus, the tolerance for position displacement is limited to about 30 $\mu m$ on the whole surface of films.

Another method based on the use of sub-MeV electron tracks from natural radioactive sources was recently proposed \cite{miyamoto}. 
With this method more accurate local alignment parameters can be assured and the position tolerance can be reduced to $\simeq10~ \mu m$. The potentialities of both methods for the purpose of reducing the combinatorial background are shown in Table \ref{tab:alignment}. 
Application to real-case neutrino events is in progress.
 
\begin{table}[htb]
\label{tab:commonway}
\begin{center}
\begin{tabular}{ccc}
\hline
Method   & X-ray mark & Sub-MeV electron\\
\hline
Position tolerance & 30 $\mu$m & 10 $\mu$m  \\
Accidental coincidence BG	& $ 0.018~ track/CSd $	& $ 0.002~ track/CSd $ \\
\hline
\end{tabular}
\end{center}
\caption {The alignment accuracy and the estimated background from accidental coincidence between two emulsion films in a CSd (125 $cm^2$). The estimation is done on the assumption 
that the track density on single films is $100~ tracks/cm^2$ and that the angular tolerance 
between two emulsion films is 10 $mrad$.}
\label{tab:alignment}
\end{table}

\subsubsection{Experimental results with X-ray fiducial marks}

\begin{figure}[htbp] 	
\centering 		
\includegraphics[width=0.45\textwidth]{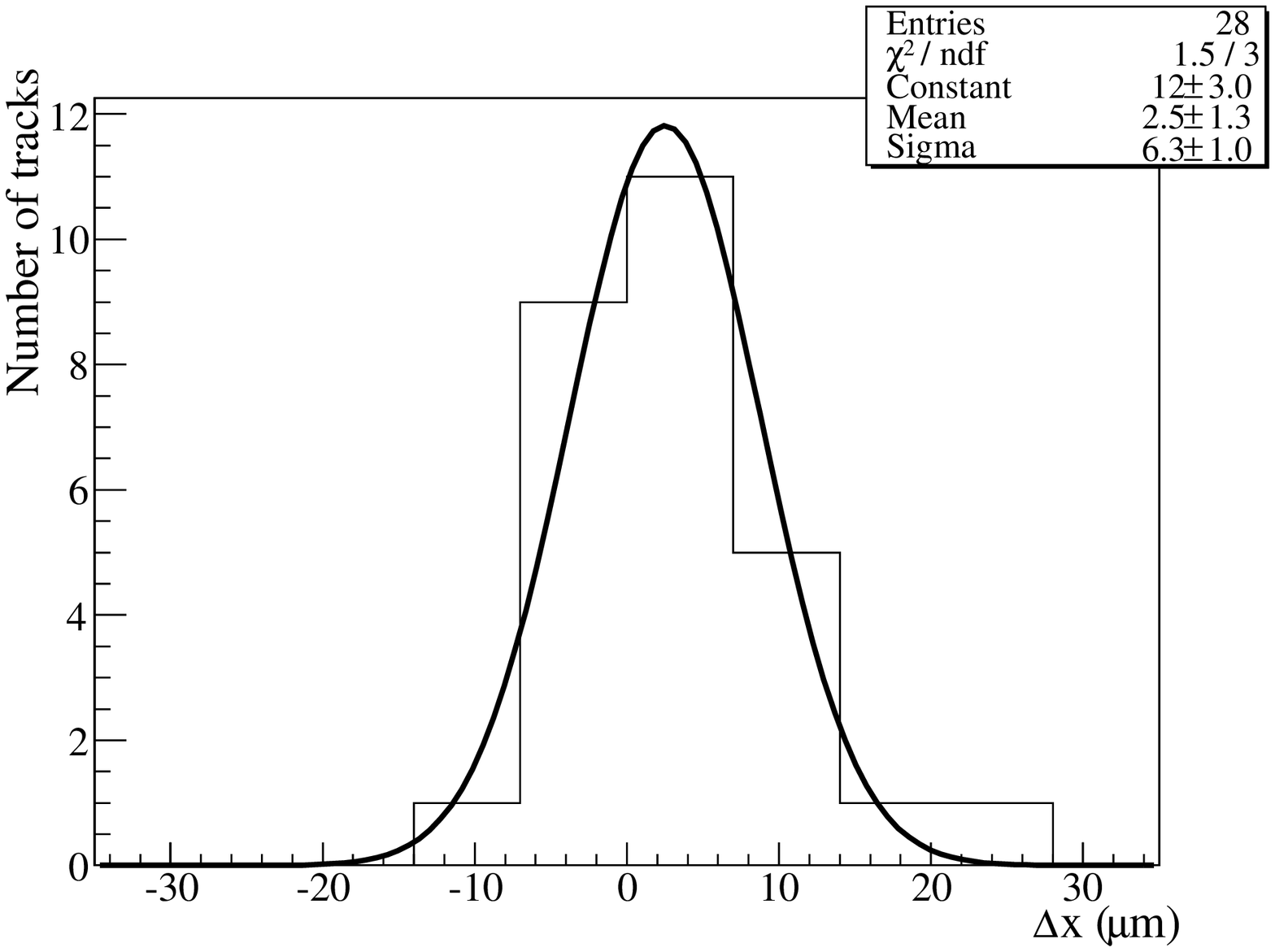} 		
\includegraphics[width=0.45\textwidth]{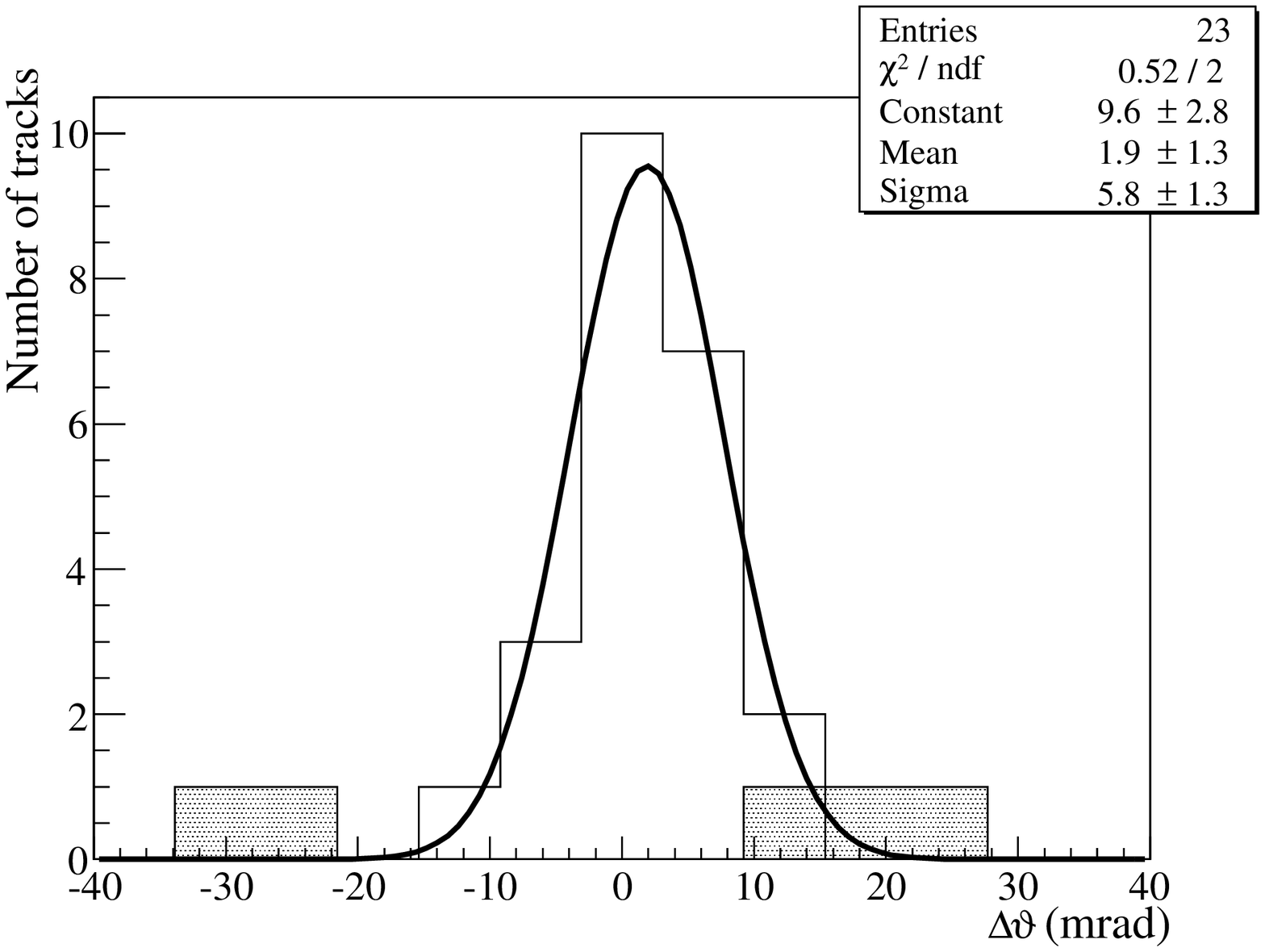} 	
\caption{ Distributions of the position and angular residuals between CSd emulsion films. For angular residuals open boxes refer to tracks formed by two base-tracks while filled boxes refer to tracks formed by a base-track and a single micro-track.}
\label{fig:dp_csd} 
\end{figure}

Results on alignment based on the X-ray mark method are shown in Fig. \ref{fig:dp_csd}. As track slopes are large the angular resolution is worse than what is anticipated for tracks from neutrino interactions. 
In the plot relative to the angular residuals the entries referring to 3-micro-track candidates are marked since the angular resolution is substantially worse for micro-tracks compared to base-tracks.
The 3-micro-track candidates contribute, as expected, to the distribution tails. 

\begin{figure}[htbp] 	
\centering
\includegraphics[width=0.45\textwidth]{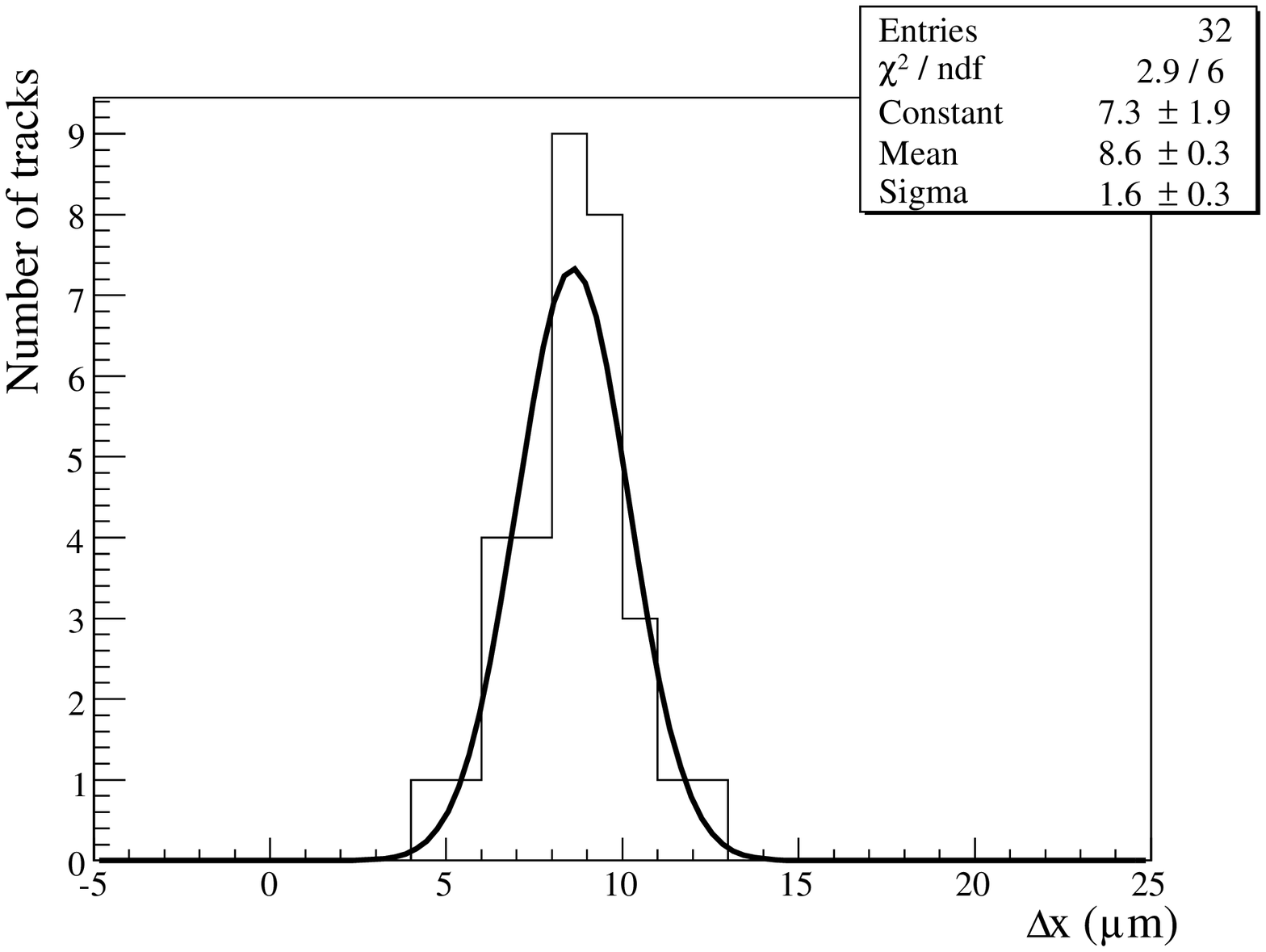} 		
\includegraphics[width=0.45\textwidth]{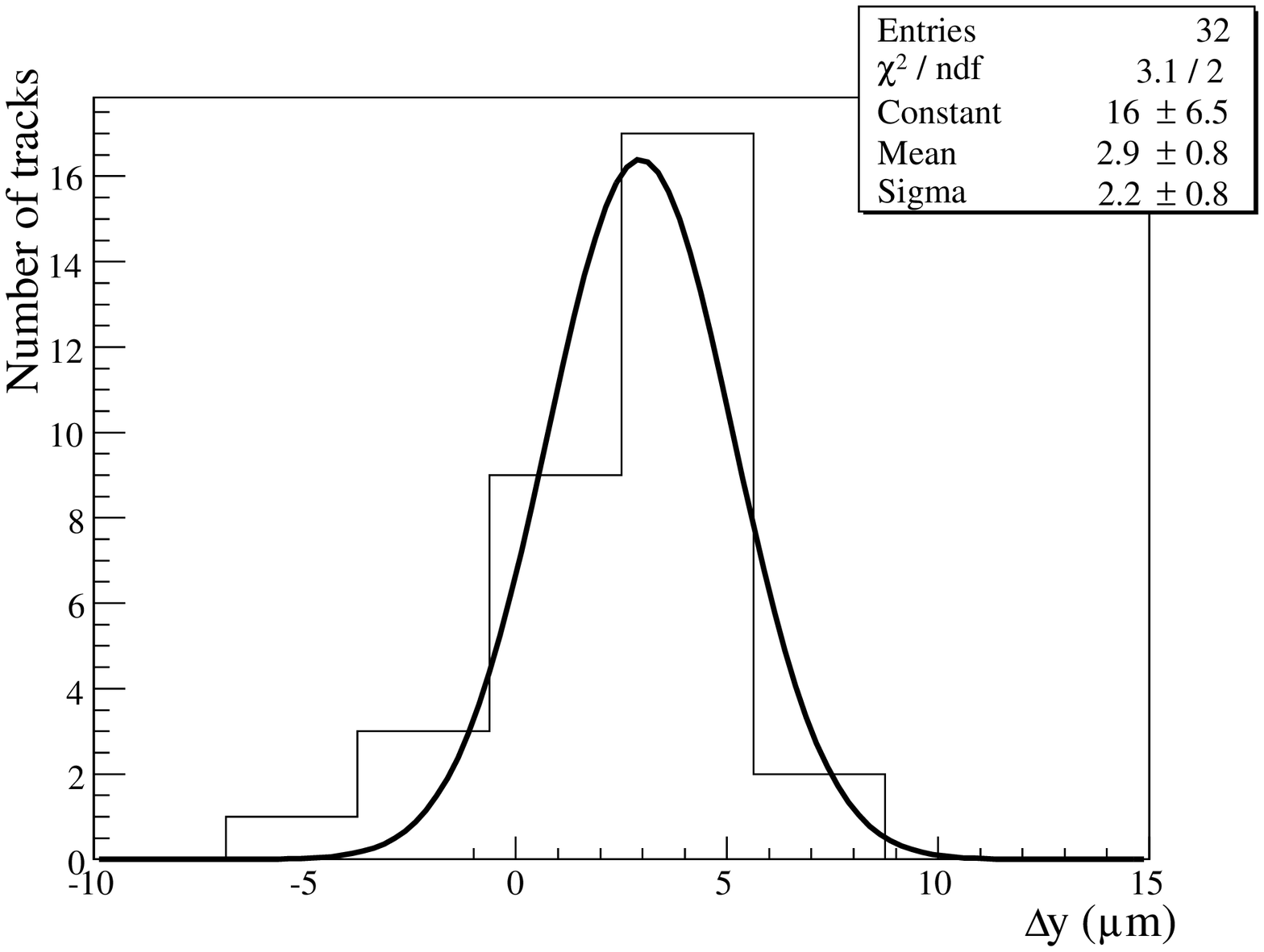}
\caption{\small Distributions of the position residuals between CSd emulsion films. The 
entries refer to tracks related to a single neutrino event.} 	
\label{fig:dx_cs} 
\end{figure}

The width of the position residuals distribution is attributed, as discussed above, to local deformations of the films. To illustrate this effect, data from a single, high-multiplicity neutrino interaction that occurred during the beam commissioning run from October 2007 were analyzed. As many as 34 secondary tracks were detected as CSd coincidences.
Among those tracks 32 were selected as impinging on the CSd within a rather limited area of the order of $ 1~cm^2 $. 
Fig. \ref{fig:dx_cs} shows the distributions of the position residuals between the two films of the same CSd. 
The position offset is a measurement of a local discrepancy with respect to the X-ray mark reference system while the standard deviation is approaching the emulsion intrinsic resolution.

\subsubsection{Experimental results with low energy electrons}

\begin{figure}[hbtp]
\begin{center}
\begin{tabular}{cc}
\includegraphics[width=0.6\linewidth]{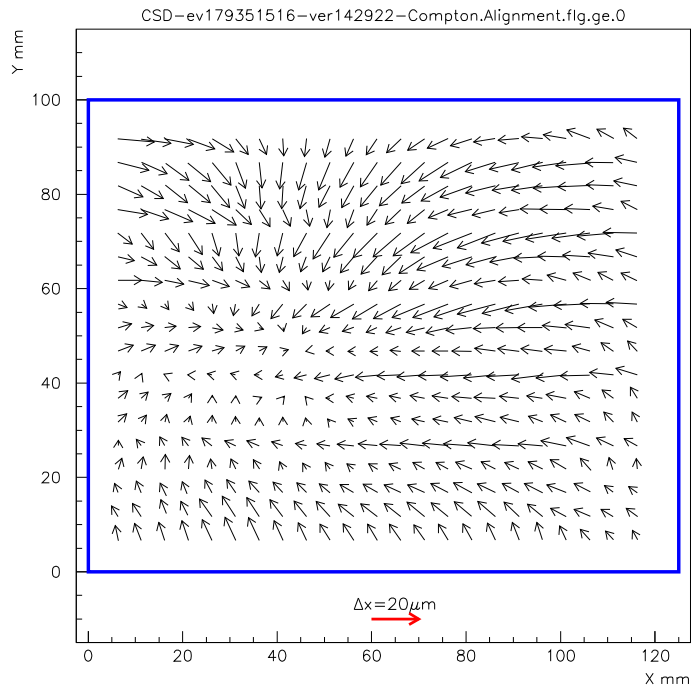}\\
\end{tabular}
\end{center}
\caption{ Local alignment parameters as a vector map for all cells. The thick line represents the edge of the CS.}
\label{fig:miyamoto}
\end{figure}

The local offset resulting from the global X-ray alignment which has been exemplified in the previous Section can be measured by using a local alignment method based on low-energy electron tracks crossing both films. 
This method was applied to all CSd extracted during the beam commissioning runs in 2007. 
A nearly full area scan of $115~mm \times 90~mm$ was performed at the Nagoya University with the S-UTS system \cite{suts}. 
The scanning area was divided into cells of $5~mm \times 5~mm$. 
The two films were first aligned by the X-ray mark and then the local offsets were calculated for each cell with low-energy electrons. 
Fig. \ref{fig:miyamoto} shows the global map of the offsets on one CSd. 
The arrows and their gradients demonstrate a relative deformation of the films induced by the difference of the strain before and after the processing and by the stress due to the setting of films on the microscope stages.

\begin{figure}[hbtp]
\begin{center}
\includegraphics[width=0.6\linewidth]{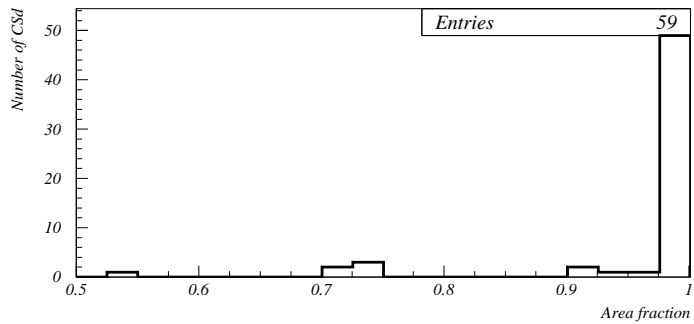}
\caption { The distribution of the film surface fraction successfully aligned locally with low energy electrons.}
 \label{fig:continual_ratio}
\end{center}
\end{figure}

The process was applied to 59 CSd. The fraction of the area where the tracks were successfully aligned is estimated for each CSd and its distribution is given in Fig. \ref{fig:continual_ratio}. The local alignment with low energy electrons is most often successful. 

\begin{figure}[bthp]
\begin{center}
\includegraphics[width=0.7\linewidth]{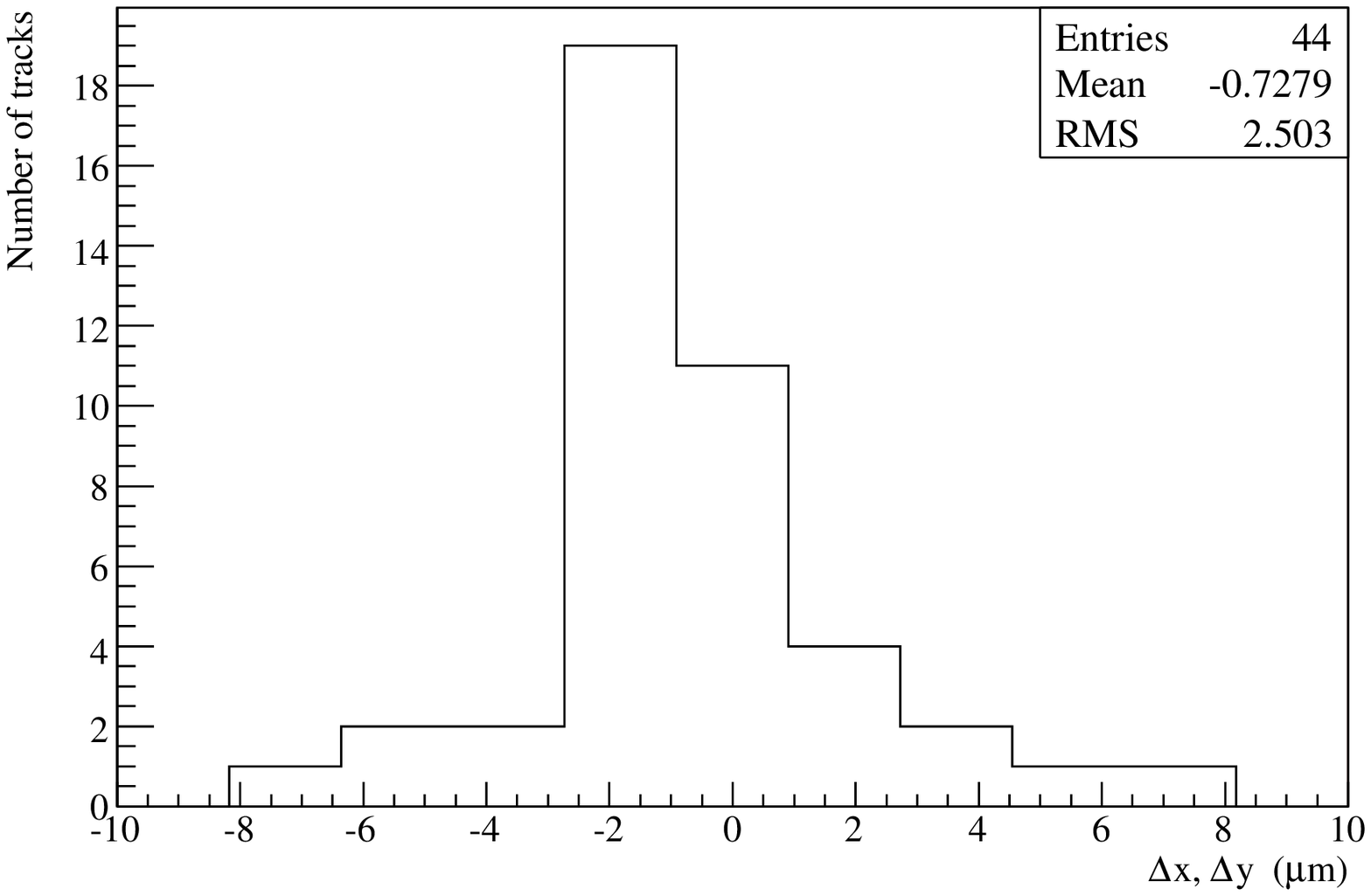}
\caption {The distribution of the position residuals between CSd emulsion films after the local offset correction. A CSd track has two entries referring to the residuals for the X and Y projections.}
\label{fig:yoshida}
\end{center}
\end{figure}

Fig. \ref{fig:yoshida} shows the distribution of the position residuals between emulsion films after the correction for the local offsets. A clear improvement in the position displacement is obtained compared to the X-ray mark method. 

\subsection{Connection between CSd and ECC brick}
\begin{figure} 	
\centering 	
		\includegraphics[width=0.45\textwidth]{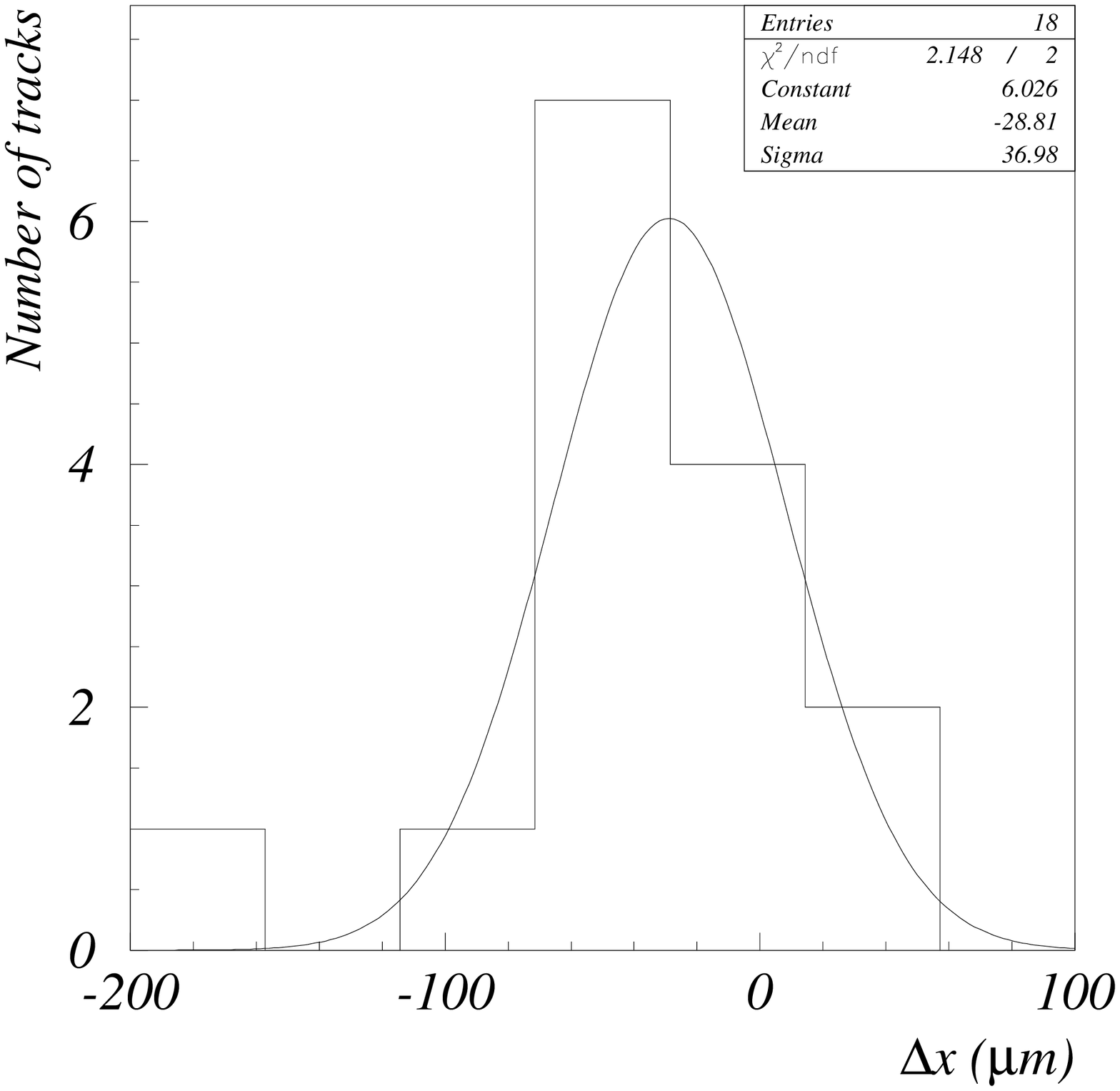}
		\includegraphics[width=0.45\textwidth]{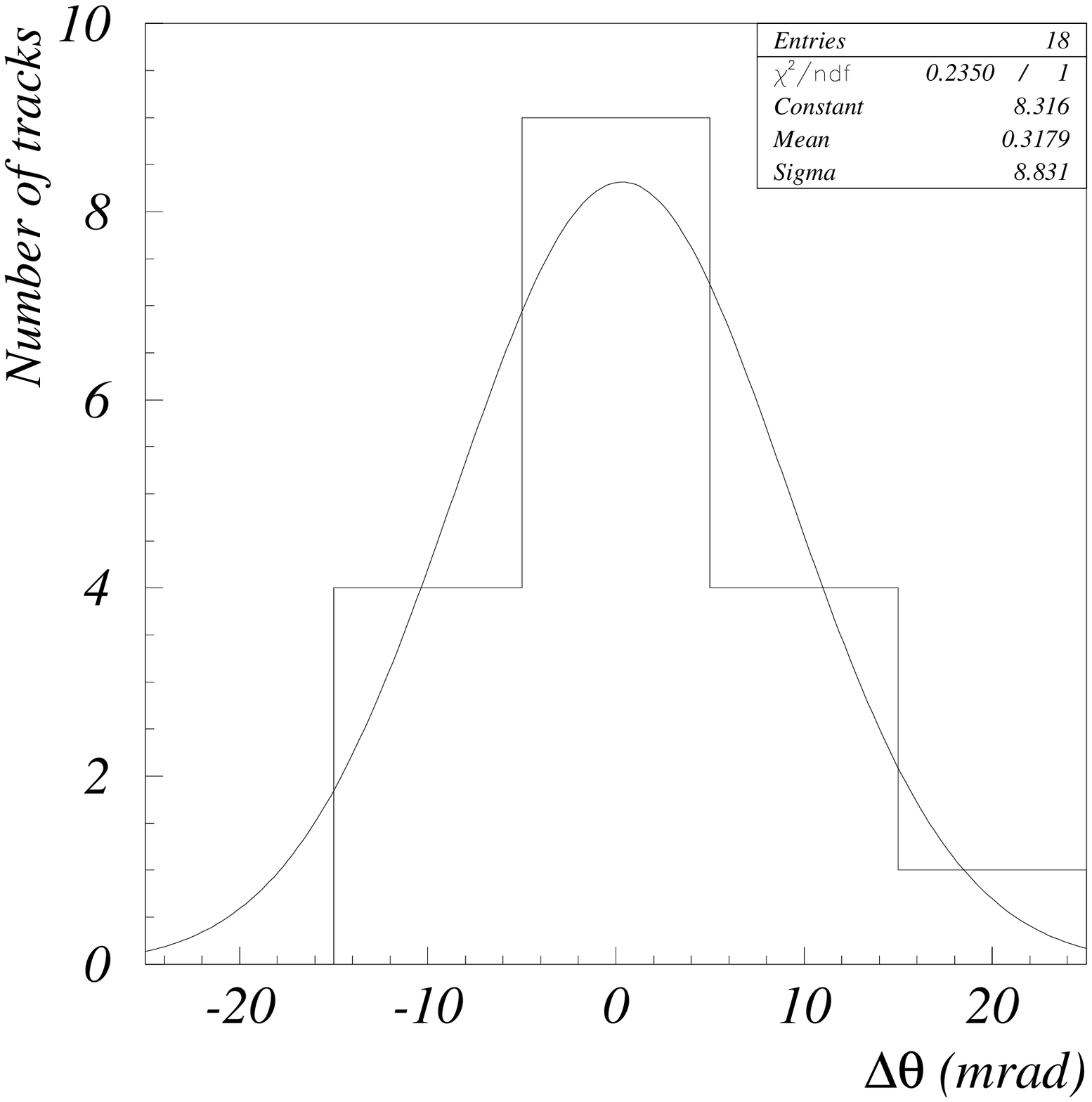}
\caption{\small Distributions of the position and angular residuals of high energy 
cosmic-ray tracks measured in the CSd and in the most downstream emulsion film of 
the ECC brick.
} 	
\label{fig:CS_to_ECC} 
\end{figure}

The X-ray marks allow aligning the CSd with the most downstream film of the corresponding ECC brick. 
An alignment accuracy of $\sim 100~ \mu m$ and of $\sim 20~mrad$ has been obtained, as shown in Fig. \ref{fig:CS_to_ECC}. Distributions of the position and angular residuals of high energy cosmic-ray tracks measured in the CSd and in the most downstream emulsion film of the ECC brick are shown.
Understanding of systematics (e.g. CSd planarity, actual distance between the CSd and the last film of a brick, etc.) that alter the alignment accuracy is in progress. More information will be provided by the on-going study of the neutrino events collected in the neutrino run of 2007.

\section{Conclusions}

An interface nuclear emulsion detector allowing for very low background detection of ionizing tracks has been realized for the purposes of the OPERA experiment. It is made of a doublet of emulsion films called CSd.

The adoption of the refreshing technique that washes away most of the latent images of cosmic-ray tracks recorded before the film storage in the underground laboratory, the accurate packing procedure, and the exploitation 
of refined alignment methods to tag genuine track coincidences, allow to keep the CSd 
backgrounds below 0.1 $tracks/100~cm^2$. 
Ultimately, starting with a global alignment based on X-ray marks of the two films of a CSd and then improving the alignment locally by tracking low energy electrons, the fake tracks background is reduced to 0.02 - 0.005 per CSd.
A production facility was constructed underground at LNGS to manufacture 154,000 CSd, 60\% of which were produced by the end of 2007. 

The commissioning of the CSd demonstrates good tracking performance in view of the neutrino interaction location and analysis for the OPERA experiment. The CSs play the central role of being the interface between the electronic detectors and the ECC bricks that constitute the target.

The partially installed OPERA detector has been commissioned with low-flux underground cosmic-rays since April 2007 and was exposed to the CNGS neutrino beam in October 2007. Triggered cosmic-ray tracks were successfully relocated in target units allowing alignment procedures to be developed and tested. The location of several neutrino interactions is in progress and will be the subject of a forthcomming publication. More details about the procedures and results presented in this paper can be found in \cite{ariga}.

Although developed for the OPERA experiment, the CS $doublet$ concept may have other applications. Indeed, a coincidence between a doublet of nuclear emulsion films with extremely low background in principle adds a sort of time resolution to emulsion detectors.  Applications may be found in accelerator and cosmic-ray experiments and in muon radiography \cite{kazan}.

\input{acknowledgement}

\include{reference}
\end{document}

%% file: author.tex
\newcommand{\ANKARA}        {1}
\newcommand{\ANNECY}        {2}
\newcommand{\ASSERGILNGS}   {3} 
\newcommand{\BARI}          {4}
\newcommand{\BARIINFN}     {5}
\newcommand{\BERN}          {6}
\newcommand{\BOLOGNAINFN}  {7}
\newcommand{\BOLOGNA}       {8}
\newcommand{\BRUSSELS}      {9}
\newcommand{\DUBNA}         {10}
\newcommand{\FRASCATI}      {11}
\newcommand{\FUNABASHI}     {12}	
\newcommand{\HAIFA}         {13}
\newcommand{\HAMBURG}       {14}
\newcommand{\GAZWADONG}     {15}
\newcommand{\KARIYA}        {16} 
\newcommand{\KOBE}          {17}
\newcommand{\LAQUILA}       {18}
\newcommand{\LYON}          {19}
\newcommand{\MOSCOWINR}     {20}
\newcommand{\MOSCOWITEP}    {21}
\newcommand{\MOSCOWLPI}     {22}
\newcommand{\MOSCOWSINP}    {23}
\newcommand{\MUNSTER}       {24}
\newcommand{\NAGOYA}        {25}
\newcommand{\NAPOLIINFN}   {26}
\newcommand{\NAPOLI}        {27}
\newcommand{\NEUCHATEL}     {28}
\newcommand{\OBNINSK}       {29}
\newcommand{\PADOVAINFN}   {30}
\newcommand{\PADOVA}        {31}
\newcommand{\ROMA}          {32}
\newcommand{\SALERNO}       {33}
\newcommand{\STRASBOURG}    {34}
\newcommand{\URBINO}        {35}
\newcommand{\UTSUNOMIYA} {36}
\newcommand{\ZAGREB}        {37}
\newcommand{\ZURICH}        {38}

\newcommand{\OperaInstitutes}{
\llap{$^{\ANKARA}$}       METU-Middle East Technical University, TR-06531 Ankara, Turkey \\
\llap{$^{\ANNECY}$}       LAPP, Universit\'e de Savoie, CNRS/IN2P3, 74941 Annecy-le-Vieux, France\\
\llap{$^{\ASSERGILNGS}$}  Laboratori Nazionali del Gran Sasso dell'INFN, 67010 Assergi (L'Aquila), Italy \\
\llap{$^{\BARI}$}         Dipartimento di Fisica dell'Universit\`a  di Bari and INFN, 70126 Bari, Italy \\
\llap{$^{\BARIINFN}$}    INFN Sezione di Bari, 70126 Bari, Italy \\
\llap{$^{\BERN}$}         University of Bern, CH-3012 Bern, Switzerland \\
\llap{$^{\BOLOGNAINFN}$} INFN Sezione di Bologna, I-40127 Bologna, Italy \\
\llap{$^{\BOLOGNA}$}      Dipartimento di Fisica dell'Universit\`a  di Bologna and INFN, I-40127 Bologna, Italy \\\llap{$^{\BRUSSELS}$}     IIHE-Inter-University Institute for High Energies, Universit\'e Libre de Bruxelles, B-1050 Brussels, Belgium \\
\llap{$^{\DUBNA}$}        JINR-Joint Institute for Nuclear Research, 141980 Dubna, Russia \\
\llap{$^{\FRASCATI}$}     Laboratori Nazionali di Frascati dell'INFN, 00044 Frascati (Roma), Italy \\
\llap{$^{\FUNABASHI}$}    Toho University, 274-8510 Funabashi, Japan \\
\llap{$^{\HAIFA}$}        Department of Physics, Technion, 32000 Haifa, Israel\\ 
\llap{$^{\HAMBURG}$}      Hamburg University, 22043 Hamburg, Germany\\
\llap{$^{\GAZWADONG}$}    Gyeongsang National University, 900 Gazwa-dong, Jinju 660-300, Korea\\
\llap{$^{\KARIYA}$}       Aichi University of Education, 448 Kariya (Aichi-Ken), Japan\\
\llap{$^{\KOBE}$}         Kobe University, 657 Kobe, Japan \\
\llap{$^{\LAQUILA}$}      Dipartimento di Fisica dell'Universit\`a  dell'Aquila and INFN, Gr. Coll. L'Aquila, Italy\\
\llap{$^{\LYON}$}         IPNL, Universit\'e Claude Bernard Lyon 1, CNRS/IN2P3, 69622 Villeurbanne, France\\
\llap{$^{\MOSCOWINR}$}    INR-Institute for Nuclear Research of the Russian Academy of Sciences, 117312 Moscow, Russia\\
\llap{$^{\MOSCOWITEP}$}   ITEP-Institute for Theoretical and Experimental Physics, 117259 Moscow, Russia \\
\llap{$^{\MOSCOWLPI}$}    LPI-Lebedev Physical Institute of the Russian Academy of Sciences, 117924 Moscow, Russia\\
\llap{$^{\MOSCOWSINP}$}   SINP MSU-Skobeltsyn Institute of Nuclear Physics of Moscow State University, 119992 Moscow, Russia \\
\llap{$^{\MUNSTER}$}      University of M\"unster, 48149 M\"unster, Germany\\
\llap{$^{\NAGOYA}$}       Nagoya University, 464-01 Nagoya, Japan\\
\llap{$^{\NAPOLIINFN}$}  INFN Sezione di Napoli, 80125 Napoli, Italy \\
\llap{$^{\NAPOLI}$}       Dipartimento di Fisica dell'Universit\`a Federico II di Napoli and INFN, 80125 Napoli, Italy \\
\llap{$^{\NEUCHATEL}$}    Universit\'e de Neuch\^atel, CH 2000 Neuch\^atel, Switzerland\\
\llap{$^{\OBNINSK}$}      Obninsk State University, Institute of Nuclear Power Engineering, 249020 Obninsk, Russia\\
\llap{$^{\PADOVAINFN}$}  INFN Sezione di Padova, 35131 Padova, Italy \\
\llap{$^{\PADOVA}$}       Dipartimento di Fisica dell'Universit\`a  di Padova and INFN, 35131 Padova, Italy \\
\llap{$^{\ROMA}$}         Dipartimento di Fisica dell'Universit\`a  di Roma ``La Sapienza" and INFN, 00185 Roma, Italy \\
\llap{$^{\SALERNO}$}      Dipartimento di Fisica dell'Universit\`a  di Salerno and INFN, 84084 Fisciano, Salerno, Italy \\
\llap{$^{\STRASBOURG}$}   IPHC, Universit\'e Louis Pasteur, CNRS/IN2P3, 67037 Strasbourg, France\\
\llap{$^{\URBINO}$}       CSAAE - Universita' di Urbino and Laboratori Nazionali di Frascati dell'INFN \\
\llap{$^{\UTSUNOMIYA}$}   Utsunomiya University, 321-8505 Utsunomiya, Japan \\
\llap{$^{\ZAGREB}$}       IRB-Rudjer Boskovic Institute, 10002 Zagreb, Croatia\\
\llap{$^{\ZURICH}$}       ETH-Eidgen\"ossische Technische Hochschulen Z\"urich, CH-8092 Zurich, Switzerland \\
\llap{$^{a}$} Now at Chonnam National University\\
\llap{$^{b}$} Now at Nihon University\\
}

\newcommand{\OperaAuthorList}{
A.~Anokhina$^{\MOSCOWSINP}$,
S.~Aoki$^{\KOBE}$,
A.~Ariga\thanks{Corresponding author.}$^{~~,~\NAGOYA}$,
L.~Arrabito$^{\LYON}$,
D.~Autiero$^{\LYON}$,
A.~Badertscher$^{\ZURICH}$,
F.~Bay$^{\ANKARA}$,
F.~Bersani~Greggio$^{\URBINO}$,
A.~Bertolin$^{\PADOVAINFN}$,
M.~Besnier$^{\ANNECY}$
% ...
\\}

%% file: acknowledgement.tex
\section {Acknowledgements}


We warmly acknowledge funding from our national agencies: 
{\it Fonds de la Recherche Scientifique - FNRS et Institut Interuniversitaire des Sciences Nucl\'eaires} for Belgium, 
MoSES for Croatia, 
IN2P3-CNRS for France, 
BMBF for Germany, 
INFN for Italy, 
the {\it Japan Society for the Promotion of Science} (JSPS), 
the {\it Ministry of Education, Culture, Sports, Science and Technology} (MEXT) and the {\it Promotion and Mutual Aid Corporation for Private Schools of Japan} for Japan, 
SNF and ETHZ for Switzerland, 
the {\it Russian Foundation for Basic Research} grants 06-02-16864-a and 08-02-91005-CERN-a for Russia,
the {\it Korea Research Foundation} (KRF-2007-313-C00161) for Korea.

We thank INFN for providing fellowships and grants to non Italian researchers, ILIAS-TARI for access to the LNGS research infrastructure and for the financial support through EU contracts P2006-01-LNGS and P2006-16-LNGS.

We are finally indebted to our technical collaborators for the excellent quality of their work over many years of design, prototyping and construction of the OPERA detector and of its facilities.